\documentclass[11pt,a4paper]{article}
\usepackage[T1]{fontenc}
\usepackage[utf8]{inputenc}
\usepackage{lmodern,microtype}
\usepackage[margin=24mm,headheight=15pt]{geometry}
\usepackage{graphicx,booktabs,tabularx,array,amsmath,enumitem,xcolor}
\usepackage{tikz,pgfplots}
\usetikzlibrary{arrows.meta,positioning,fit,backgrounds}
\pgfplotsset{compat=1.18}
\usepackage{caption,placeins,fancyhdr,url}
\definecolor{praxis}{HTML}{174F45}
\definecolor{soft}{HTML}{ECF4F0}
\definecolor{amber}{HTML}{9B651F}
\usepackage[colorlinks=true,linkcolor=praxis,citecolor=praxis,urlcolor=praxis]{hyperref}
\hypersetup{pdftitle={LearnAdapt Praxis: Controlled AI Assistance and Evidence Traces for Adult Workplace Learning},pdfauthor={Nizam Kadir}}
\setlist{nosep,leftmargin=*}
\newcolumntype{P}[1]{>{\raggedright\arraybackslash}p{#1}}
\newcolumntype{Y}{>{\raggedright\arraybackslash}X}
\newcommand{\code}[1]{\texttt{#1}}
\newcommand{\release}{\texttt{40bf713}}
\title{\LARGE\bfseries LearnAdapt Praxis: Controlled AI Assistance and Evidence Traces for Adult Workplace Learning}
\author{Nizam Kadir\\\small Science, Mathematics and Technology\\\small Singapore University of Technology and Design, Singapore\\\small\href{mailto:nizam_kadir@mymail.sutd.edu.sg}{nizam\_kadir@mymail.sutd.edu.sg}}
\date{28 September 2026}
\begin{document}
\maketitle\thispagestyle{fancy}
\begin{abstract}
AI can help adults produce plausible workplace artefacts while leaving their independent reasoning difficult to inspect. LearnAdapt Praxis addresses this design problem through a five-stage learning workspace: Frame, Learn, Build, Validate and Transfer. The application stores an initial response, versioned specifications, coaching records, validation observations, a separate transfer response and facilitator feedback. Project-level server controls restrict coaching, access to earlier work and learner export during an active transfer attempt; they do not establish that a learner avoided assistance outside the application. This technical report describes the implemented architecture and evaluates selected workflow, authority, failure-recovery and provenance properties. A reproducible synthetic rehearsal exercised 120 serial project episodes across three fictional workplace contexts and four injected provider conditions. All 3,990 recorded checks met their specified expectations. The resulting records contained 240 specification versions, 1,200 activity events and 60 stored synthetic hints; 60 injected provider failures produced no fabricated hints. Separate authentication regressions exposed and resolved a session-expiry defect, and limited staging and production checks verified account access and live model connectivity. The evidence supports the tested engineering properties of the recorded release. It does not establish learning gains, model-output quality, unaided assessment validity, population usability or production capacity. The contribution is an implemented arrangement for separating assisted work, project-level assistance withdrawal and inspectable evidence, accompanied by executable tests and an explicit account of what remains to be validated with adult learners.
\end{abstract}

\section{Introduction}
A useful AI-generated workplace artefact is not, by itself, evidence that its user can explain or reproduce the reasoning behind it. This distinction matters when an adult learning programme asks participants to design an AI-supported service, test its limits and take responsibility for its use. A polished specification may conceal unexamined assumptions about source reliability, affected users, escalation or the division of responsibility between a model and a person. Conversely, a learner who identifies a limitation or rejects an attractive automation may demonstrate valuable judgement without producing an impressive prototype.

Praxis is a web application in the LearnAdapt ecosystem that makes these decisions visible within a structured project. Learners begin with their own starting response, engage with course material, develop a five-field specification, record validation observations and respond to a different scenario. Facilitators configure learning conditions and review the resulting evidence. The application is available at \url{https://www.learnadaptresearch.org/praxis/}. Its public pages, authentication, learner workspace and facilitator workspace form one application; the public, device-local demonstration is a separate illustrative mode.

This report asks three engineering research questions:
\begin{enumerate}
\item \textbf{RQ1:} Can the implementation preserve the selected workflow and role boundaries while allowing an authorised project to reach review and export?
\item \textbf{RQ2:} Does it retain an inspectable record of revisions, assistance, independent-task status and review, including when a provider fails?
\item \textbf{RQ3:} What operational evidence supports a bounded demonstration or pilot, and what remains outside the tested scope?
\end{enumerate}
The study contributes a system description, a frozen executable rehearsal and a release-specific verification record. It presents no human-participant experiment and no new learning-effect estimate. Its unit of evaluation is a scripted project episode, with individual assertions nested within episodes.

\section{Design rationale and related work}
\subsection{Assistance, feedback and independent performance}
Pea's account of scaffolding makes the withdrawal of support central to distinguishing scaffolded learning from continuing performance with distributed resources~\cite{pea}. Praxis implements a coarse withdrawal point: the learner begins an independent transfer attempt and the service changes what that learner can request for that project. This is a design response to the distinction, not a validated adaptive fading algorithm. The system does not estimate a learner's latent competence or decide that assistance should fade on the basis of a psychometric model.

Formative-feedback research motivates feedback that helps a learner act on a task rather than merely receive a score~\cite{shute}. Praxis accordingly asks its coach for a concise diagnostic question and limited scaffold, while human facilitators supply review judgements. These are prompt instructions and interface choices. The present tests verify that the intended instructions and context reach the provider callback; they do not establish that a deployed model consistently follows them.

Research on productive failure provides a reason to preserve initial attempts and subsequent reasoning, while also showing that the sequence and consolidation of learning activities matter~\cite{kapur}. Praxis's locked starting response is not an implementation or replication of that intervention. It preserves one trace that a future study could interpret alongside instruction, revision and transfer.

A randomised high-school mathematics study by Bastani and colleagues distinguished assisted practice from later unassisted performance and found that the consequences of generative AI depended on the tutoring arrangement~\cite{bastani}. This provides a motivation to measure independent outcomes separately. Its population, tasks and intervention differ from adult workplace design; its findings are not treated as an expected effect size for Praxis.

\subsection{Relationship to LearnAdapt and Critsly}
LearnAdapt Agentic Studio and PedOS describe a broader programme of configurable educational AI and governed execution~\cite{pedos}. Praxis is a specific application within that programme. The implementation examined here uses shared authentication infrastructure with a dedicated Praxis router, domain service, user interface and deployment target. This report does not claim that Praxis currently integrates every LearnAdapt application, exchanges learner models across products or provides an operating system for all pedagogy.

The author's Critsly/StudioCrit technical report used synthetic rehearsals to distinguish operational readiness from educational effectiveness~\cite{studiocrit}. Praxis adopts that reporting distinction for a different workflow: adult workplace problem framing, design decisions, validation and transfer. No Critsly account counts, classifications, timings or simulated learner observations are reused as Praxis results. The new rehearsal includes its input-generating script, per-check outcomes, exported records and source fingerprint, so that the stated counts can be recomputed.

\section{Application and system architecture}
\subsection{Deployment and authority boundaries}
\begin{figure}[htbp]
\centering
\begin{tikzpicture}[node distance=9mm,box/.style={draw=praxis,rounded corners,fill=soft,align=center,text width=4cm,minimum height=13mm,font=\small},arr/.style={-{Stealth},thick,draw=praxis}]
\node[box] (browser) {Browser\\Learner / facilitator UI};
\node[box,right=13mm of browser] (entry) {Canonical route\\\code{/praxis/}\\Praxis service};
\node[box,below=of entry] (router) {Praxis router\\Session, role and CSRF checks};
\node[box,below=of router] (domain) {Praxis domain service\\Access and workflow\\Revision checks};
\node[box,left=13mm of domain] (db) {Database\\Projects, cohorts, sessions};
\node[box,below=of domain] (ai) {AI provider adapter\\Context and coaching request};
\draw[arr] (browser)--(entry);\draw[arr] (entry)--(router);\draw[arr] (router)--(domain);\draw[arr,<->] (domain)--(db);\draw[arr,<->] (domain)--(ai);
\node[align=left,text width=4.2cm,font=\small,above=6mm of db] {Returned state is filtered for the current actor and project phase.};
\end{tikzpicture}
\caption{Implemented request and persistence boundaries. The rehearsal substitutes an injected callback for the AI adapter and uses SQLite in memory; separate live checks exercise the deployed service. The diagram is a description of the implementation, not a claim of independently audited security.}
\label{fig:architecture}
\end{figure}
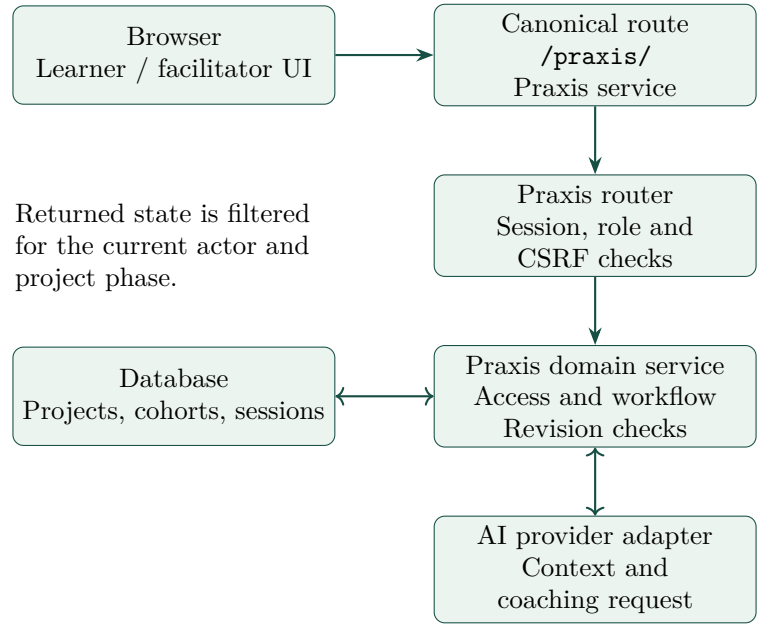

The canonical website mounts Praxis at \code{/praxis/}. Requests are proxied to an isolated PHP/Apache Cloud Run service; the live service uses a shared database for application and session persistence. The dedicated session cookie is scoped to the Praxis path. Authentication, authorisation and cross-site request forgery (CSRF) checks precede protected operations. The browser submits commands and renders returned state; authoritative workflow checks execute in the domain service (Figure~\ref{fig:architecture}).

Each project has an owner and may belong to a cohort. The cohort's owner can inspect and review associated projects but cannot use learner commands to overwrite their submissions. An unrelated learner or facilitator cannot read the project through the tested service methods. These restrictions are more specific than a general educator role: access also depends on the relationship between the actor, cohort and project.

A project stores a course-configuration snapshot at creation, including its challenge, learning content, coaching policy, transfer scenario and configuration version. Later cohort edits affect new projects. Existing records therefore retain the conditions with which they began. This is useful for interpretation, although the stored policy identifier is not a complete snapshot of model infrastructure, provider configuration or all historical prompts.

\subsection{Five stages and a separate transfer state}
Table~\ref{tab:workflow} relates the visible stages to stored evidence. The specification is a structured learning artefact, not an executable workplace application. A learner may link a prototype and describe tests, but Praxis does not execute those external tests or certify deployment readiness.

\begin{table}[htbp]\centering\small
\caption{Learning workflow and the limits of its evidence.}\label{tab:workflow}
\begin{tabularx}{\linewidth}{P{18mm}YY}\toprule
Stage & Implemented record or rule & Interpretation limit \\\midrule
Frame & Starting response is saved and locked before coaching. & An authenticated submission does not prove authorship or absence of outside help. \\
Learn & Course material and learner reflection are retained. & Saving a reflection does not establish understanding. \\
Build & Five specification fields; numbered versions; hint records. & A specification is not a running application or a validated design. \\
Validate & Observations and optional prototype URL are recorded. & Text describes a claimed test; the system does not verify that it occurred. \\
Transfer & New scenario; prior content and coaching restricted for the learner's active project. & Restriction does not cover other applications, prior downloads or all other projects. \\
Review & Assigned facilitator records rubric judgements and feedback. & The rubric has not undergone reliability or validity testing in this report. \\\bottomrule
\end{tabularx}
\end{table}

The server checks for a locked baseline before later substantive commands. It requires a saved specification and validation observations before transfer begins. It does not impose a fully linear instructional sequence: for example, a saved reflection is not a server prerequisite for saving a specification. The learning interface proposes a sequence, while the server enforces a smaller set of explicit conditions.

During active transfer, the learner-facing project response removes the baseline, reflection, specification fields, prior versions, hints and validation observations; only the transfer scenario is retained from the course configuration. Learner export and coaching are rejected. The assigned facilitator may still inspect the project. A sufficiently long, self-attested response completes the attempt and restores the learner's earlier record. An abandonment route clears the active response and records that event. Figure~\ref{fig:state} shows the principal boundary; abandonment was inspected in source but was not a condition in the 120-episode rehearsal.

\begin{figure}[htbp]\centering
\begin{tikzpicture}[box/.style={draw=praxis,rounded corners,align=center,text width=39mm,minimum height=17mm,font=\small},arr/.style={-{Stealth},draw=praxis,thick}]
\node[box,fill=soft] (learn) {Learning phase\\Baseline and versions\\Coaching available};
\node[box,fill=soft,right=9mm of learn] (transfer) {Active transfer\\Learner access restricted};
\node[box,fill=soft,right=9mm of transfer] (done) {Submitted transfer\\Earlier evidence restored};
\draw[arr] (learn)--node[above,font=\scriptsize]{start}(transfer);\draw[arr] (transfer)--node[above,font=\scriptsize]{submit}(done);
\draw[arr] (transfer.south) -- ++(0,-8mm) -| node[below,pos=.25,font=\scriptsize]{abandon; event retained}(learn.south);
\node[align=center,text width=125mm,font=\small,below=18mm of transfer] {Project-level controls apply to this service. They cannot establish that an adult avoided previously viewed material, other tools or other people.};
\end{tikzpicture}
\caption{Principal assistance boundary. Transfer submission requires a minimum response length and a self-attestation. Neither condition measures competence or verifies unaided authorship.}
\label{fig:state}
\end{figure}

\subsection{Coaching and recoverable provider failure}
The coaching request includes the current scenario, reflection, specification, the last four hint replies and a learner question. Its instruction asks for one concise question and at most one scaffold, discourages complete submissions and prohibits grading or claims that tests ran. Learner artefacts are explicitly described to the model as untrusted data. No experimental claim about prompt-injection resistance is made here.

Before calling the provider, the service writes a pending-request marker with a nonce and expiry. It uses an optimistic revision check to reject stale updates. While a request is pending, a conflicting save is rejected. A returned response is saved only when the pending nonce still matches. Successful records include reply, question, provider, model, reported token count and policy identifier. An exception or empty reply clears the marker, records unavailability and preserves the saved specification. This failure path offers a reload-and-retry message instead of substituting a fabricated response.

These controls manage application state around a provider call. They do not enforce the semantic quality, word count or pedagogical appropriateness of every model response. Nor does the reentrant test below establish correctness under every possible multi-process race or provider timeout.

\begin{figure}[htbp]\centering
\includegraphics[width=\linewidth]{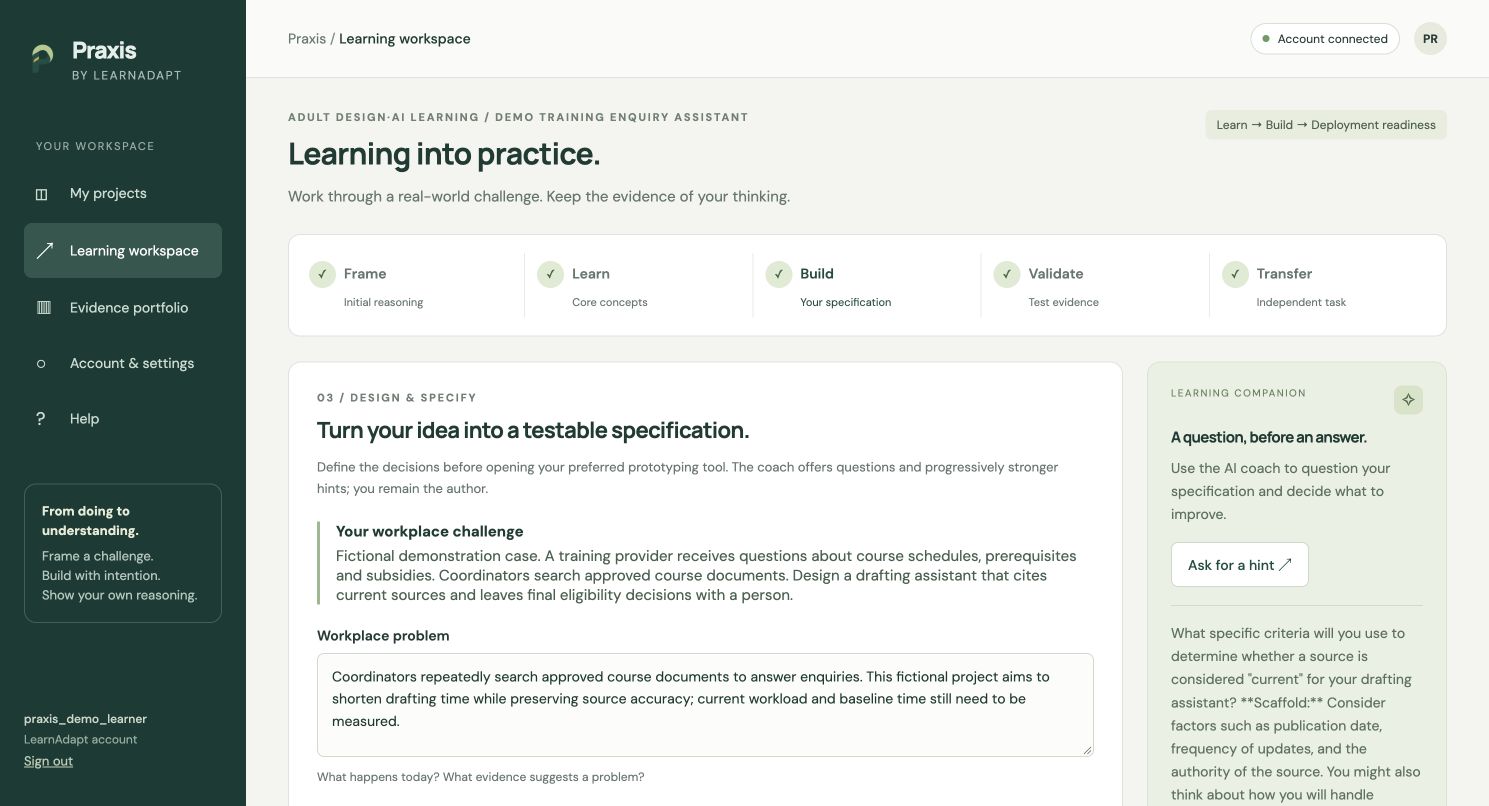}
\caption{Real production Build screen, captured on 26 September 2026. The fictional training-enquiry project contains a saved specification and a previously generated live hint. This screenshot illustrates the interface; it is not one of the 120 synthetic domain-service episodes.}
\label{fig:build}
\end{figure}
\FloatBarrier

\section{Technical study methods}
\subsection{Design, release and units of analysis}
The rehearsal was run on 28 September 2026 against the unchanged \code{PraxisService.php} from release \release. The complete Git identifier and SHA-256 fingerprint are provided in Appendix~\ref{app:repro}. A protocol was written before execution and retained with the study bundle; it was not externally preregistered. The checks were informed by the implementation and executed with AI assistance. They constitute developer-led verification rather than independent testing.

The design crossed three fictional workplace contexts with four provider conditions and ten repetitions per cell: $3\times4\times10=120$ episodes (Table~\ref{tab:design}). Each episode used a distinct synthetic learner identifier, joined a configured cohort and created one project. Two additional identities represented an unrelated learner and an unrelated facilitator. An assigned facilitator role supplied scripted review values. No person produced a study response or evaluated a rubric.

The contexts varied task text, not a simulated cognitive ability distribution. Repetitions exercised the same specified transitions with different project identities. They should not be interpreted as 120 statistically independent learners, a representative sample of adults or a naturalistic simulation of workplace behaviour. The 3,990 checks are nested assertions, with different numbers per provider condition; they are not 3,990 independent trials.

\begin{table}[htbp]\centering\small
\caption{Rehearsal conditions. Every cell contains ten scripted episodes, for 40 per context and 30 per provider condition.}\label{tab:design}
\begin{tabularx}{\linewidth}{Yrrrr}\toprule
Fictional context & Valid & Exception & Empty & Reentrant \\\midrule
Training enquiries & 10 & 10 & 10 & 10 \\
Facilities handovers & 10 & 10 & 10 & 10 \\
SME knowledge work & 10 & 10 & 10 & 10 \\\midrule
Total & 30 & 30 & 30 & 30 \\\bottomrule
\end{tabularx}
\vspace{2mm}\parbox{\linewidth}{\footnotesize Valid: deterministic non-empty reply. Exception: injected runtime exception. Empty: empty content. Reentrant: attempted overlapping save from inside the provider callback, followed by a valid deterministic reply. These are test doubles, not four language models.}
\end{table}

\subsection{Episode procedure and outcome checks}
Each episode attempted a complete authorised path: create project; lock baseline; save reflection; save two specification versions; request coaching; record validation; start transfer; save and submit a transfer response; record facilitator feedback; export the project; and read the saved state through a second service instance. The two specification versions differed in an explicit escalation responsibility. All response text was scripted and labelled synthetic.

Negative probes were interleaved with that path. They requested another user's record, attempted facilitator edits and learner self-review, tried coaching before baseline, attempted a baseline overwrite and stale-revision save, and tested coaching, export and earlier-work edits during transfer. Transfer submission was also attempted with too little text and without attestation. A rejection passed only when the expected domain exception occurred; an unexpected success failed the check. Positive checks compared returned state, counts, provenance fields and the final exported state with specified expectations.

For coaching, a callback counter established whether a denied request reached the provider abstraction. Exception and empty-response conditions tested preservation and recovery; reentrancy tested a save attempt after the pending marker had been persisted. The callback also checked that the coaching instruction and specification context were present. It returned a fixed synthetic response when successful, so no automated or human judgement of model content entered the 3,990-check total.

All outcomes were retained in \code{checks.jsonl}. \code{episodes.csv} records each episode's context, provider condition, check count, failures, provider-call count, hint count, version count, event count, final revision and elapsed time. One complete synthetic JSON export was retained per episode. No failed episode or assertion was excluded.

\subsection{Execution environment and timing}
The domain rehearsal ran serially on an arm64 macOS host using PHP 8.5.4 and SQLite 3.51.3 in memory. Each episode's duration used a monotonic clock and included its service calls, assertions and writing its exported JSON file. It excluded HTTP, session authentication, browser rendering, network transport and real model inference. Consequently, these durations are reproducibility diagnostics, not an estimate of application latency or Cloud Run capacity.

The production service uses PHP 8.2 and a different database environment. Its separate smoke checks therefore complement the local rehearsal but do not remove environment differences. There was no concurrent-user load test, controlled hardware benchmark, prolonged soak test or browser-compatibility sample.

\subsection{Authentication and deployed checks}
A separate disposable, file-backed SQLite harness exercised the real Praxis router, authentication middleware and database session handler at both the root path and \code{/praxis}. It altered only test-session timestamps to trigger idle, absolute and storage expiry. A JavaScript harness exercised the actual public authentication script with controlled network outcomes, including wrong passwords, unrelated errors, refresh failure and bounded recovery. These checks are reported separately from the project rehearsal.

Live verification used the existing demonstration roles on staging and production. Each environment received one clearly identified temporary synthetic project to exercise the baseline-to-transfer path and an actual provider response. The temporary projects were removed after verification; the three prepared production demonstration cases retained their original identifiers and revisions. Browser checks also confirmed a stale-tab recovery on staging and learner and facilitator access on production. No password-recovery email was sent, so email delivery is outside the verified scope.

\section{Results}
\subsection{Workflow, authority and evidence integrity}
All 120 episodes completed the scripted path. All 3,990 recorded checks met their expected outcomes, with zero unexpected failures. Table~\ref{tab:checks} groups the checks by the property they exercised; grouping reflects the prewritten script, not post hoc learner coding.

\begin{table}[htbp]\centering\small
\caption{Recorded checks in the 120-episode domain rehearsal. Passing means that the specified assertion or expected rejection occurred; counts do not estimate population reliability.}\label{tab:checks}
\begin{tabularx}{\linewidth}{Yrr}\toprule
Property group & Checks & Unexpected failures \\\midrule
Actor and project authority & 840 & 0 \\
Baseline and workflow prerequisites & 360 & 0 \\
Revision conflict and pending-request handling & 150 & 0 \\
Provider context, result and recovery & 720 & 0 \\
Evidence and event provenance & 720 & 0 \\
Transfer restrictions and submission & 1,080 & 0 \\
State read by a second service instance & 120 & 0 \\\midrule
Total & 3,990 & 0 \\\bottomrule
\end{tabularx}
\end{table}

The final exports contained 240 specification versions and 1,200 events: two versions and ten events per episode. The event sequence included project creation, baseline, reflection, two version saves, the coaching outcome, validation, transfer start, transfer submission and review. Draft-saving actions do not all append an event, so event counts should not be equated with a complete record of every interaction or keystroke.

All 120 active-transfer responses passed the composite redaction check. All 120 active-transfer coaching requests were rejected without calling the injected provider; all 120 learner exports during that phase were rejected. Earlier material was available again after submission. The facilitator retained inspection access while the learner's project was restricted. These observations support the selected project-level boundary in RQ1; they do not show that a participant could not obtain help through another route.

Every final export retained the submitted/attested transfer state, server-assigned reviewer identity, source-labelled timestamped events and notices separating activity records from learning-gain inference and research participation. Reopening the same database through a second service instance reproduced the saved project state in all 120 episodes. This tests a second read path on the same in-memory database, not persistence through process termination or database recovery.

\begin{figure}[htbp]\centering
\includegraphics[width=\linewidth]{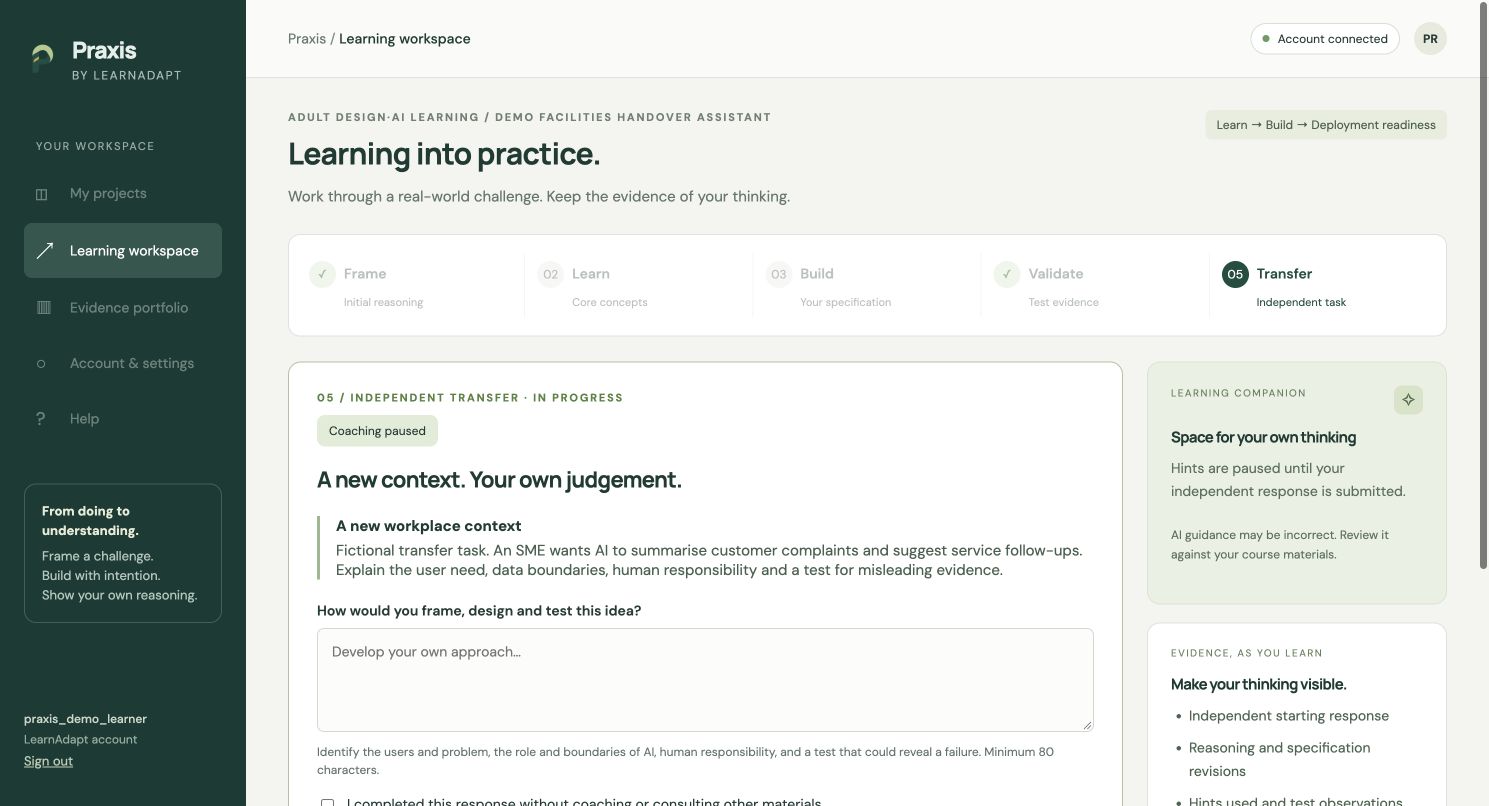}
\caption{Real production screen for the fictional facilities case with transfer active, captured on 26 September 2026. The learner's earlier project content and coach are restricted. The screen documents an application state, not verified independent human performance.}
\label{fig:transfer}
\end{figure}

\subsection{Provider failures and overlapping changes}
\begin{figure}[htbp]\centering
\begin{tikzpicture}
\begin{axis}[width=.94\linewidth,height=60mm,ybar stacked,bar width=16mm,ymin=0,ymax=36,ytick={0,10,20,30},ylabel={Project episodes},symbolic x coords={Valid,Exception,Empty,Reentrant},xtick=data,enlarge x limits=.2,legend style={at={(.5,1.02)},anchor=south,legend columns=2,draw=none,font=\small},tick label style={font=\small},grid=major,major grid style={gray!15}]
\addplot[fill=praxis,draw=praxis] coordinates {(Valid,30)(Exception,0)(Empty,0)(Reentrant,30)};
\addplot[fill=amber!65,draw=amber] coordinates {(Valid,0)(Exception,30)(Empty,30)(Reentrant,0)};
\legend{Stored synthetic hint,Logged unavailability}
\end{axis}
\end{tikzpicture}
\caption{Observed coaching outcomes by injected provider condition. Each bar represents 30 scripted episodes. Both outcomes are correct for their assigned condition. No real-model quality score is represented.}
\label{fig:provider}
\end{figure}
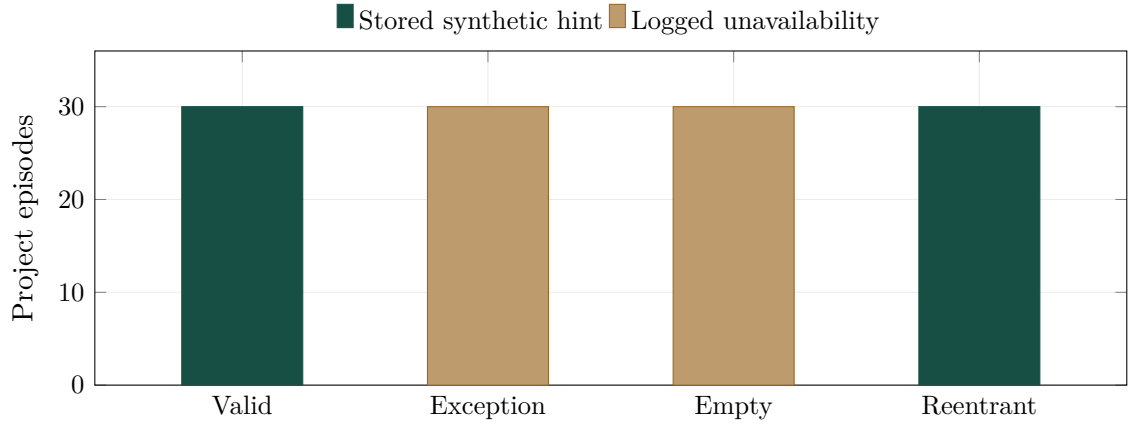

The valid and reentrant conditions each stored 30 synthetic hints. The exception and empty conditions each recorded 30 unavailable outcomes and stored zero hints (Figure~\ref{fig:provider}). Across all conditions, 120 callbacks were invoked and 60 hints were retained. The absence of hints in failure conditions was the intended outcome; it was not missing data or an unsuccessful learner episode.

All provider conditions retained the saved build fields and cleared the pending marker. All 30 reentrant save attempts were rejected while the callback was pending. Stale-revision writes were rejected once in every episode. The test does not exhaust arbitrary interleavings, shared-session races or distributed database failures, so RQ2 is supported only for the exercised state transitions.

Diagnostic episode times ranged from 0.856 to 1.620 ms, with a median of 0.923 ms and nearest-rank 95th percentile of 1.091 ms. These small values reflect serial in-memory domain operations and a test double. They should not be compared with user-facing AI response times or interpreted as evidence for supporting a particular class size.

\subsection{Session-expiry defect and deployed verification}
The authentication harness reproduced a concrete defect in the preceding release. When timeout enforcement destroyed an authenticated session, the same request could render a new login form without persisting a new session for its token. The subsequent login POST then failed CSRF validation. A second failure mode arose when an older login tab retained a token after another tab signed out.

The recorded release establishes a fresh anonymous session after timeout. A public authentication request explicitly rejected for token expiry can obtain a new uncached, same-origin token and retry once. Wrong passwords and unrelated errors are not retried automatically. Protected workspace writes from an expired session return an authentication-required response before performing work. This correction preserves passwords, roles, CSRF validation and the existing 30-minute idle and four-hour absolute limits.

The final regressions passed 58 HTTP checks, 37 JavaScript assertions and 23 existing service checks (Table~\ref{tab:verification}). These are separate denominators and are not added to the 3,990 rehearsal checks. A staging browser test confirmed successful submission from an unchanged old login tab after sign-out in another tab. The tested container image was promoted by its immutable digest to the isolated Praxis production service.

\begin{table}[htbp]\centering\small
\caption{Release-specific verification outside the 120-episode rehearsal, completed on 28 September 2026.}\label{tab:verification}
\begin{tabularx}{\linewidth}{P{32mm}P{23mm}Y}\toprule
Layer & Result & Scope \\\midrule
Router/session harness & 58 checks passed & Disposable database; two URL mounts; expiry and token rejection/recovery. \\
Authentication UI harness & 37 assertions passed & Actual public script; controlled responses; bounded retry and error handling. \\
Existing domain tests & 23 checks passed & Previously defined ownership, workflow, review and invitation checks. \\
Staging smoke & 27 API checks passed & Both roles; temporary synthetic project; one live OpenAI response. \\
Production smoke & 37 API checks passed & Both roles; three existing cases; temporary project; one live OpenAI response; original cases unchanged. \\\bottomrule
\end{tabularx}
\end{table}

Both live coaching responses reported OpenAI and \code{gpt-4o-mini}; reported token counts were 419 on staging and 428 on production. These single responses establish successful invocation and persistence at the recorded time. They are not an evaluation sample for response accuracy, learning support, provider stability or cost forecasting. Public-route checks also confirmed the canonical mount, public pages, protected application entry and static assets including the favicon.

\begin{figure}[htbp]\centering
\includegraphics[width=\linewidth]{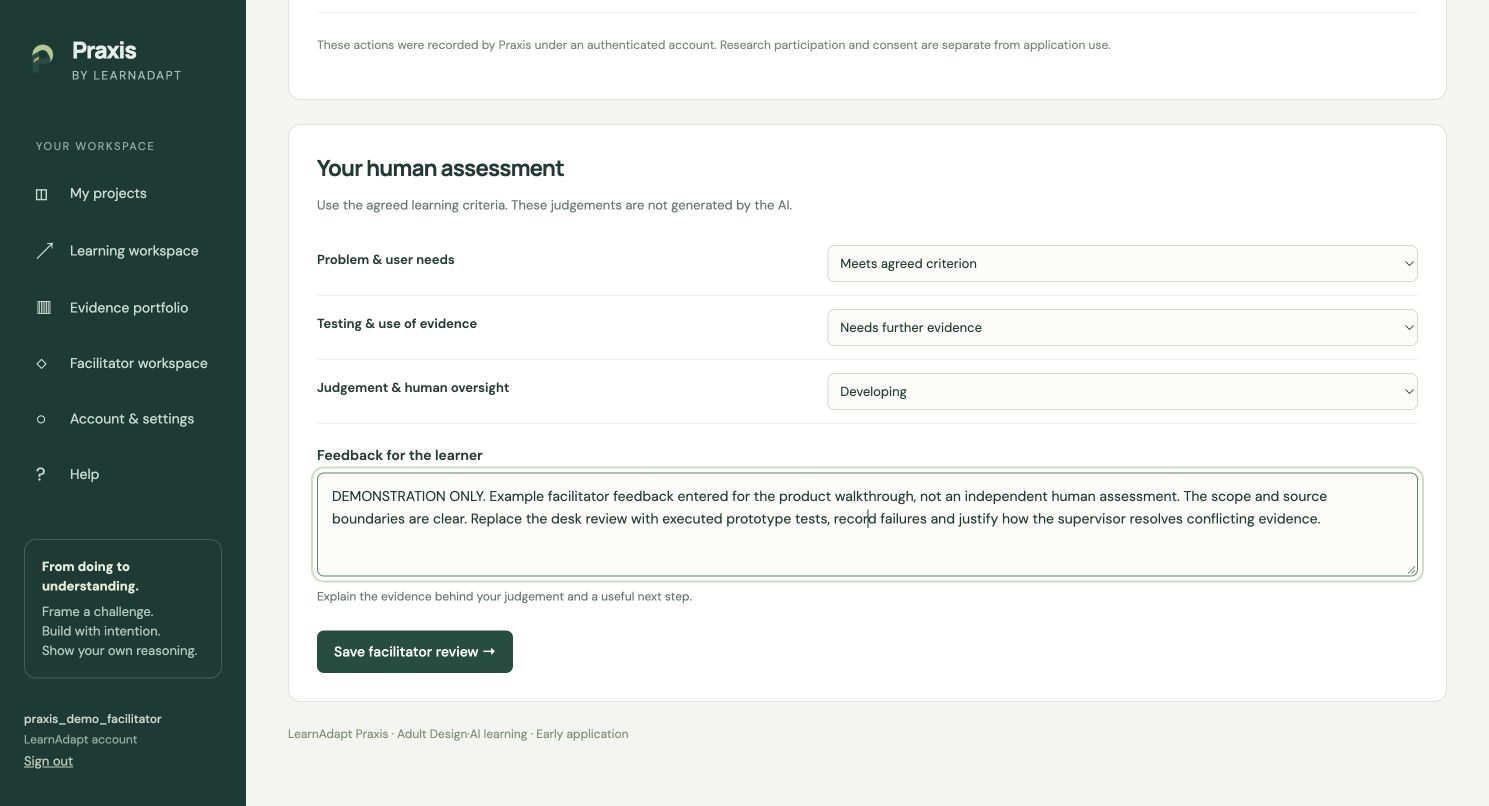}
\caption{Real production facilitator assessment controls, captured on 26 September 2026. The example feedback is visibly labelled demonstration material. The selected rubric values illustrate the application; they are not independent human ratings used in this study.}
\label{fig:review}
\end{figure}
\FloatBarrier

\section{Discussion}
\subsection{An inspectable separation of assistance and evidence}
The strongest supported result is that the implemented service can preserve a selected set of boundaries while retaining a complete, reviewable project record through the scripted workflow. The learner's initial response remains distinguishable from later versions, coaching remains attributable to a provider record, and facilitator feedback remains distinguishable from a model reply. This arrangement can make future studies more interpretable by retaining the conditions under which an artefact was produced.

The contribution is a concrete system arrangement and its executable verification. It is not a new language model, a validated measure of adult competence or evidence that this interface is superior to an existing learning environment. The all-pass result means that the specified tests passed against the frozen implementation. It does not demonstrate the absence of bugs, especially because the tests were written with knowledge of that implementation.

\subsection{Provenance does not establish educational validity}
W3C PROV distinguishes the entities, activities and agents involved in producing information~\cite{prov}. Praxis retains useful local analogues: project versions, activity events and authenticated actors. The present JSON schema is not represented as a conformant PROV serialisation. It also does not capture every interaction, store a tamper-evident audit log or preserve a full model-request archive. A provider/model label and policy identifier are helpful but insufficient for exact reconstruction of a historical model response.

Table~\ref{tab:evidence} makes the inference boundary explicit. In particular, an event label such as authenticated learning activity describes an application provenance category, not a psychological finding. A scripted user and a real adult can generate records with the same schema; their evidential status depends on the study design and collection context.

\begin{table}[htbp]\centering\small
\caption{What an exported record supports, and what requires additional evidence.}\label{tab:evidence}
\begin{tabularx}{\linewidth}{P{30mm}YY}\toprule
Record element & Supported interpretation & Unsupported inference \\\midrule
Baseline and versions & Text was saved in a particular project sequence. & The change measures a learning gain. \\
Hint metadata & A response and its reported origin were stored. & The advice was accurate, safe or beneficial. \\
Validation observations & An actor recorded a description of testing. & The external prototype passed those tests. \\
Transfer attestation & The actor asserted independent completion. & Assistance outside Praxis was absent. \\
Facilitator rubric & An authorised reviewer supplied values. & Ratings are reliable or psychometrically calibrated. \\
Events and revisions & Recorded transitions and version checks are inspectable. & The log is complete, immutable or sufficient for causal attribution. \\\bottomrule
\end{tabularx}
\end{table}

\subsection{Limits of the independent-task boundary}
The transfer controls apply to the current project and the tested API paths. A learner may already have seen or downloaded earlier content, remember a hint, consult another project or use an external tool. Browser navigation restrictions cannot remove those possibilities. The system also permits an assigned facilitator to inspect records during transfer, and its review route is not gated on completion of the transfer. A course protocol must therefore determine when and how facilitator feedback is delivered if unaided performance is an outcome of interest.

The starting and transfer scenarios are not established as equivalent forms. They differ in surface context and may differ in difficulty, familiarity or consequence. Subtracting a baseline rubric score from a transfer score would not automatically produce a valid learning-gain measure. Future work needs task design, counterbalancing where appropriate, assessment criteria and independent scoring before such comparisons become interpretable.

\subsection{Readiness, scalability and governance}
The production checks demonstrate that the selected workflows could be exercised through the canonical service with both demonstration roles. They support a bounded demonstration and inform pilot preparation. They do not establish service-level availability, concurrent-user capacity, accessibility conformance, robustness on diverse devices or recovery from infrastructure failures. Registration and invitation handling have existing local tests, but live email-based recovery was not tested in this verification.

The service records course and project information associated with user identifiers. Research use requires its own consent and governance arrangements; application use alone is not consent to research. Before a human study, the programme needs decisions about approved learning materials, model-provider data handling, role assignment, retention, deletion, participant information and the separation of identity mappings from analytical exports. No claim of completed ethics approval is made for a future participant study in this report.

The session-expiry finding also illustrates why demonstration readiness requires more than a fresh-login smoke test. An account can remain valid while the path from an expired session to a usable login form fails. Reproducing that lifecycle and retaining the regression is more informative than resetting the account password and assuming the underlying problem has disappeared.

\section{Future empirical evaluation}
A future adult-learning study should first examine whether the workflow is understandable and usable in the intended course. Observations should distinguish difficulties with the workplace problem from difficulties caused by the interface. Facilitator preparation, the duration of activities and acceptable forms of AI help should be specified rather than inferred from application events.

A subsequent evaluation could compare course-defined assistance conditions while holding learning objectives, task access and instructional time as constant as practicable. Independent performance should be assessed with clearly defined, preferably supervised tasks and a prespecified rubric. Delayed transfer may be useful when the research question concerns retained workplace reasoning. Raters should be trained and agreement reported before reconciliation; automated traces should remain explanatory process measures rather than substitute outcome scores.

The technical protocol should also expand to actual multi-process conflict tests, provider interruption and timeout behaviour, long-running sessions, network loss, account recovery and broader device/accessibility checks. Model evaluation should use a separate, diverse set of adult-learning prompts with prespecified criteria for answer-giving, misleading advice, source uncertainty and harmful overconfidence. No sample size, power result, effectiveness claim or commercially ready status follows from the current rehearsal.

\section{Conclusion}
LearnAdapt Praxis implements a structured adult workplace-learning project with controlled access to AI assistance, versioned artefacts, a project-level transfer boundary and facilitator review. The frozen synthetic rehearsal completed 120 episodes and 3,990 specified checks, while targeted authentication tests identified and resolved a real session-expiry defect. Limited live verification established functioning account access, saved evidence and model connectivity on the recorded release. These results support the exercised engineering properties and provide a reproducible starting point for further testing. Whether the arrangement improves adults' independent judgement, transfer or workplace outcomes remains an empirical question.

\section*{Declarations}
\textbf{Study scope and ethics.} The reported evaluation uses scripted synthetic content, test identities and demonstration accounts. It contains no recruited participants, participant outcome data or independent human rubric ratings. Human-participant research and any associated ethics approval are future work.

\textbf{Author contribution and interests.} The report concerns the author's LearnAdapt application and research programme. The author has a direct interest in the system's development; the evaluation is developer-led and is not an independent certification. No funded NRF trial or grant outcome is reported here.

\textbf{AI assistance.} OpenAI Codex assisted with code inspection and repair, protocol and test-script preparation, execution of synthetic checks, result aggregation, reference verification and manuscript preparation. Synthetic responses and facilitator-role feedback are scripted test fixtures. Actual model replies were used only for the separate live connectivity checks described in the report. The author is responsible for the final interpretation and submission.

\textbf{Data and software availability.} The companion study bundle contains the frozen domain service, executable rehearsal, protocol, 3,990 check records, 120 episode summaries, 120 synthetic project exports, authentication regression harnesses, source fingerprints and analysis code. The manuscript source includes original production screenshots and their provenance information. The bundle is prepared for distribution with this report; no public archival DOI is claimed. The live application is not a substitute for the frozen source and recorded evaluation artefacts. Credentials and private deployment secrets are excluded.

\appendix
\section{Reproduction and audit trail}\label{app:repro}
The study is tied to application commit \path{40bf7134f5c73b72d671f67517954e1c8d3e07f9}. The frozen domain service has SHA-256 fingerprint:
\begin{quote}\small\path{44bae32e94a1602859f286ab2b5644eab093122818f4b35381657e8d3b473c07}\end{quote}
From the companion \code{Study} directory, run:
\begin{verbatim}
php run_rehearsal.php rerun-results
python3 analyse.py rerun-results
python3 authentication/tests/praxis_auth_http_test.py
node authentication/tests/praxis_auth_ui_test.mjs
php authentication/tests/praxis_service_test.php
\end{verbatim}
PHP requires PDO SQLite. The authentication HTTP harness starts a temporary localhost server, creates disposable credentials and a file-backed test database, and removes its temporary directory when finished. The JavaScript checks require Node.js; the analysis uses the Python standard library. No live credentials or model API key are required for these local checks. A separate output directory preserves the original results.

Random project identifiers, timestamps and diagnostic timings will differ on a rerun. Under the frozen protocol, the expected aggregate counts are 120 episodes, 3,990 checks, 120 provider callback invocations, 60 hints, 240 specification versions and 1,200 events. These expectations are design-specific, not targets against which results should be edited. Unexpected failures should be retained and investigated.

\begin{table}[htbp]\centering\small
\caption{Companion artefacts and their audit purpose.}\label{tab:artifacts}
\begin{tabularx}{\linewidth}{P{54mm}Y}\toprule
Artefact & Purpose \\\midrule
\code{PROTOCOL.md} & Pre-execution design and interpretation boundaries. \\
\code{source/PraxisService.php} & Frozen domain implementation used by the rehearsal. \\
\code{run\_rehearsal.php} & Exact synthetic inputs, callback conditions and assertions. \\
\code{results/checks.jsonl} & Per-check category and pass/fail outcome. \\
\code{results/episodes.csv} & Per-episode denominator, outcome counts and timings. \\
\code{results/export-*.json} & Complete synthetic final records for all episodes. \\
\code{environment.json} & Runtime details, release identity and fingerprints. \\
\code{analyse.py} & Recomputes aggregates from retained raw outputs. \\
\code{authentication/} & Frozen router, middleware and regression harnesses. \\\bottomrule
\end{tabularx}
\end{table}

\section{Screenshot provenance}
Figures~\ref{fig:build}, \ref{fig:transfer} and \ref{fig:review} are actual browser captures of the production application at the canonical \code{/praxis/} route on 26 September 2026. They show deliberately fictional demonstration content. Original PNG files and a capture manifest are retained with the manuscript source. The September 28 release changes session recovery; it does not alter the depicted learning workflow. Screenshots are interface evidence and are analysed separately from the synthetic rehearsal outputs. No screenshot is presented as a captured participant session or a measured educational outcome.
\end{document}